\documentclass[aip,apl,preprint]{revtex4-1}
\usepackage{graphicx}
\usepackage{dcolumn}
\usepackage{bm}

\begin{document}

\preprint{}

\title{Quantitative determination of anisotropic magnetoelectric coupling in BiFeO$_{3}$-CoFe$_{2}$O$_{4}$ nanostructures}

\author{Yoon Seok Oh}
\affiliation{CeNSCMR, Department of Physics and Astronomy, Seoul National University, Seoul 151-747, Republic of Korea}
\author{S. Crane}
\affiliation{Department of Materials Science and Engineering, University of California, Berkeley, CA 94720, USA}
\author{H. Zheng}
\affiliation{Department of Materials Science and Engineering, University of California, Berkeley, CA 94720, USA}
\author{Y. H. Chu}
\affiliation{Department of Materials Science and Engineering, National Chiao Tung University, HsinChu 30010, Taiwan}
\author{R. Ramesh}
\affiliation{Department of Materials Science and Engineering, University of California, Berkeley, CA 94720, USA}
\author{Kee Hoon Kim}
\affiliation{CeNSCMR, Department of Physics and Astronomy, Seoul National University, Seoul 151-747, Republic of Korea}

\date{\today}

\begin{abstract}
The transverse and longitudinal magnetoelectric susceptibilities (MES) were quantitatively determined for (001) heteroepitaxial BiFeO$_{3}$-CoFe$_{2}$O$_{4}$ nanostructures. Both of these MES values were sharply enhanced at magnetic fields below 6 kOe and revealed asymmetric lineshapes with respect to the dc magnetic field, demonstrating the strain-induced magnetoelectric effect. The maximum transverse MES, which reached as high as $\sim$60 mV/cm Oe, was about five times larger than the longitudinal MES. This observation signifies that transverse magnetostriction of the CoFe$_{2}$O$_{4}$ nanopillars is enhanced more than the bulk value due to preferred magnetic domain alignment along the [001] direction coming from compressive, heteroepitaxial strain.
\end{abstract}

\maketitle

The magnetoelectric (ME) effect is a physical phenomenon in which the electric polarization $P$ (magnetization $M$) is modulated by the magnetic field $H$ (electric field $E$). There is a growing interest in the application of ME effects toward various devices, including magnetic sensors~\cite{nanjap} and energy harvesters.~\cite{dong} As such, numerous efforts have been made to obtain strong ME couplings in novel ME composites made of ferroelectric (or piezoelectric) and ferromagnetic materials.~\cite{boomgaard,jryu} In these types of ME composites, $P$ is varied with $M$ via the strain ($u$)-coupling at the interface between the piezoelectric and magnetostrictive phases. Thus, the configuration of this interface is a significant control parameter for determining the extent of ME coupling.~\cite{nanjap} In this respect, a layered sandwich structure [i.e., (2-2) structures] produces a larger ME coupling than for particulates that are dispersed in a matrix [i.e., (0-3) structures] because the former has a larger interface area. In contrast, this common practice cannot be applied to multilayered thin films, where clamping of the nonmagnetic substrate can prevent strain-coupling between the layers.~\cite{nanprl}

In 2004, an epitaxial thin film composed of CoFe$_{2}$O$_{4}$ (CFO) nanopillars embedded in a BaTiO$_{3}$ (BTO) matrix [i.e., (1-3) structures] was grown and suggested to be an alternative to circumvent the substrate clamping effect. On the other hand, it has been quite difficult to determine quantitatively the ME coupling of such nanostructures and, more generally, numerous ME films, except observing the existence of nontrivial ME coupling.~\cite{zhengscience}  One dominant reason for this difficulty is that the ME voltage signal is proportional to the film thickness; it typically becomes smaller than 1 $\mu$V for film thicknesses less than 1 $\mu$m. To overcome this difficulty, a large ac magnetic field ($H_{\rm{ac}}$) of up to $\sim$1 kOe was recently used to obtain the effective MES as a function of $H_{\rm{ac}}$.~\cite{yanapl,yanjap} However, direct measurements of the MES based on a conventional scheme that employs small $H_{\rm{ac}}$ values with variations in the dc magnetic field ($H_{\rm{dc}}$) would be useful for understanding the ME coupling of numerous multiferroic films or nanostructures at the quantitative level. In this work, by use of the conventional scheme, we provide experimental evidences that a 300 nm thick BiFeO$_{3}$-CoFe$_{2}$O$_{4}$ (BFO-CFO) nanostructure has a peculiar MES anisotropy that is not expected in bulk forms of those materials.

The self-assembled epitaxial BFO-CFO film with thickness of 300 nm was grown on a (001) SrRuO$_{3}$/SrTiO$_{3}$ substrate by pulsed laser deposition.~\cite{zhengadvm} The film had the CFO nanopillars embedded in a BFO matrix with a volume fraction of 1:1. Top electrodes Pt/SrRuO$_{3}$ were deposited on the film surface (Fig. 1(a)). For magnetic hysteresis measurements, a vibrating sample magnetometer was utilized. For ferroelectric hysteresis loops, the displacement current was measured using a fast digitizer and a high voltage amplifier.

\begin{figure}
\begin{center}
\includegraphics[width=0.48\textwidth]{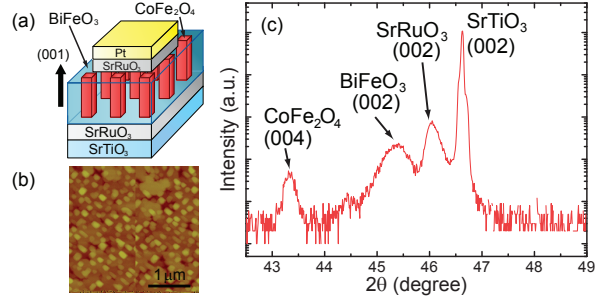}
\end{center}
\caption{ (a) Schematic picture, (b) AFM image, and (c) XRD pattern of the BFO-CFO film with nanopillar structures.}
\label{fig1}
\end{figure}

To investigate the MES (${\alpha}{\equiv}{\delta}P/{\delta}H_{\rm{ac}}$), especially for thin films, we developed a highly sensitive ME susceptometer that operates inside the PPMS (Quantum Design). A pair of solenoids was designed to induce a $H_{\rm{ac}}$ of $\sim$4 Oe inside the solenoid pair and a voltage pick-up coil was used to determine the phase of $H_{\rm{ac}}$. In particular, modulated charges, instead of voltages, were measured using a high-impedance charge amplifier with a gain factor of $10^{12}$ V/C. This makes the amplified signal independent of film thickness but proportional to electrode area so that the signal-to-noise ratio was improved. Based on this scheme, we were able to detect small ME charges in a thin film with a thickness $\sim$40 nm and a circular electrode of diameter $\sim$100 $\mu$m. The lowest charge noise ($\Delta Q$) was ${\sim}10^{-17}$ C, which corresponded to a voltage noise of $\Delta Q/C_{\rm{s}}$ ($C_{\rm{s}}$ = sample capacitance). The determined $\alpha$ could be converted into the MES expressed as a voltage unit $\alpha _{\rm{E}}{=}{\delta}E/{\delta}H_{\rm{ac}}$ using the relationship ${\alpha}{=}{\varepsilon}{\alpha}_{\rm{E}}$, where $\varepsilon$ is the absolute permittivity of a specimen.

\begin{figure}
\begin{center}
\includegraphics[width=0.48\textwidth]{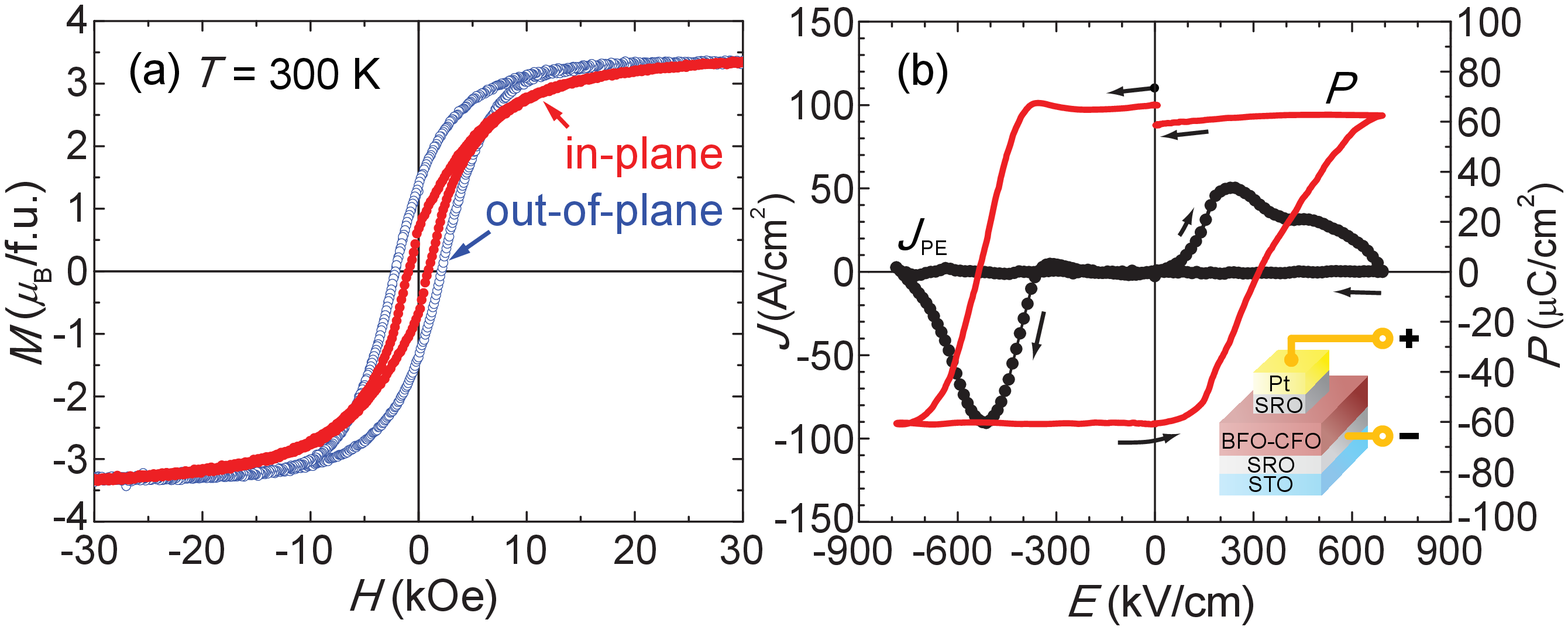}
\end{center}
\caption{(a) Magnetic hysteresis loops for the in-plane and out-of-plane directions measured at 300 K. $M$ is normalized to the volume fraction of CFO. (b) $J_{\rm{PE}}(E)$ (solid circle) was integrated to estimate the $P$-$E$ hysteresis loop (solid line).}
\label{fig2}
\end{figure}

Fig. 1(b) shows an atomic force microscopy (AFM) image of the BFO-CFO film, in which the CFO phase appears as rectangles embedded in the BFO matrix.~\cite{zhengadvm} Fig. 1(c) further shows an x-ray diffraction (XRD) pattern of the film obtained through a $\theta -2\theta$ scan around the (002) SrTiO$_{3}$ peak. The distinct (00$l$) peaks of CFO, BFO, and SrRuO$_{3}$ are consistent with the previous result~\cite{zhengadvm} that each phase was epitaxially grown on the SrTiO$_{3}$ substrate. The $d$-spacing of the CFO nanopillars was estimated as 2.0867 $\rm{{\AA}}$ from the (004) peak. This indicates a compressive strain along the [001] direction, $u_{001}{=}-0.33\%$, compared with the bulk CFO. Fig. 2(a) shows the magnetic hysteresis loops measured along the in-plane (i.e., $H{\parallel}[100]$) and out-of-plane (i.e., $H{\parallel}[001]$) directions. The saturated moment of $\sim$3.4 $\mu _{B}$/f.u. for both directions was in good agreement with the reported CFO value.~\cite{pauthenet} Moreover, there existed a large uniaxial magnetic anisotropy with an easy axis along the [001] direction. A linear extrapolation of the in-plane loop yielded a magnetic anisotropy field of $\sim$25 kOe. In a previous study on BTO-CFO nanostructures, the compressive strain of CFO caused by heteroepitaxial growth was found to be a primary contribution to the uniaxial magnetic anisotropy;~\cite{zhengapl} a large magnetic anisotropy field of 51 kOe was observed for $u_{001}{=}-1.1\%$, while the anisotropy field decreased for smaller $u_{001}$. Therefore, even in the BFO-CFO film studied here, the magnetic anisotropy field of $\sim$25 kOe seems to originate from the presence of a compressive, heteroepitaxial strain of $u_{001}{=}-0.33\%$ inside the CFO nanopillars.

To determine the $P$-$E$ loop, displacement current $J(E)$ was measured while negative, positive, and zero biases were applied successively in sequence (i.e., $-E$ to $E$ to 0). The $J(E)$ curve showed two extremes at $-500$ kV/cm and 200 kV/cm, at which a reversal of $P$ occurred. In addition to these extremes, a nonlinear background was found in the $J(E)$ curve. This could have been from either a Schottky barrier at the interface or a ferroelectric diode effect.~\cite{choi,yang} After subtracting the nonlinear background, the remaining current density $J_{\rm{PE}}(E)$ was integrated. In the obtained $P$-$E$ loop (Fig. 2(b)), the polarization value was normalized by the volume fraction of BFO. The saturated $P$ ($P_{\rm{s}}$) of $\sim$62 $\mu$C/cm$^{2}$ is comparable to the previously obtained value in an epitaxial (001) BFO film.~\cite{wang} After fully poling the sample along the positive $P$ direction, i.e. top electrode direction as indicated in the inset of Fig. 2(b), all the MES measurements were subsequently performed.

\begin{figure}
\begin{center}
\includegraphics[width=0.48\textwidth]{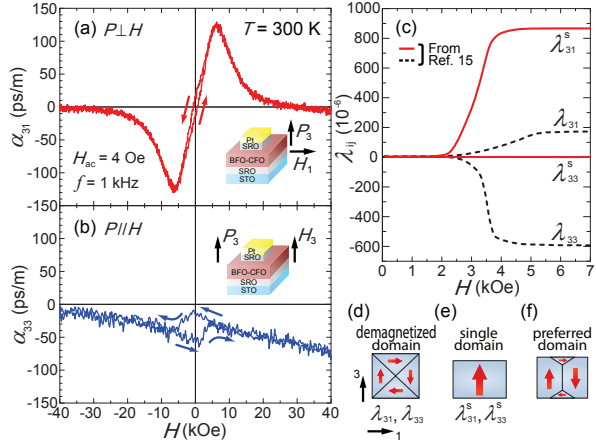}
\end{center}
\caption{(a) Transverse ($\alpha_{31}$) and (b) longitudinal MES ($\alpha_{33}$) of the BFO-CFO nanostructure at 300 K. (c) Dashed (solid) lines represent the transverse ($\lambda_{31}$) and longitudinal ($\lambda_{33}$) magnetostriction curves of a Co$_{0.8}$Fe$_{2.2}$O$_{4}$ crystal with the demagnetized (single) magnetic domain. (d) Demagnetized, (e) single, and (f) preferred magnetic domain patterns of CFO are schematically drawn.} \label{fig3}
\end{figure}

Fig. 3 summarizes transverse ($\alpha _{31}$) and longitudinal ($\alpha _{33}$) MES curves as a function of $H_{\rm{dc}}$. We note that the similar $\alpha _{31}$ and $\alpha _{33}$ curves were obtained at many different electrode spots of the same film and at those of the same kind of nanostructured film with a thickness $\sim$40 nm. The $\alpha _{31}$ clearly shows a sign reversal with the direction of $H_{\rm{dc}}$ and develops extreme points at $H_{\rm{dc}}{=}{\mp}6$ kOe. This is the archetypal lineshape expected in the strain-coupled ME media composed of piezoelectric and magnetostrictive materials, supporting that the measured MES data are reliable.
In addition, the $\alpha_{33}$ curve exhibited an asymmetric lineshape with $H_{\rm{dc}}$, which was similar to the case of $\alpha_{31}$. However, in this $\alpha _{33}$ curve, a small but non-negligible offset at $H_{\rm{dc}}{=}0$ was observed, of which value was proportional to the electrode area. Thus, this offset was attributed to a small contribution from eddy currents generated inside the electrode due to ${\delta}H_{\rm{ac}}$. Except for this offset, the $\alpha _{33}$ curve was almost an odd function of $H_{\rm{dc}}$. This further supports that the strain-coupling is a dominant source of the longitudinal ME effect as well.

As shown in Fig. 3, $\alpha _{31}{>}0$ for $H_{\rm{dc}}{>}0$ while $\alpha _{33}{<}0$ for $H_{\rm{dc}}{>}0$. This experimental result reflects that $P_{3}$ (i.e., $P{\parallel}[001]$) increases under $H_{\rm{dc}}{\parallel}[100]$, while $P_{3}$ decreases under $H_{\rm{dc}}{\parallel}[001]$. When the $H_{\rm{dc}}{\parallel}[100]$ was applied to the CFO crystal, the transverse magnetostriction ($\lambda _{31}$) was positive (dashed lines in Fig. 3(c)). Thus, the CFO nanopillars and the BFO matrix, via strain-coupling, were expected to be elongated along the [001] direction.~\cite{bozorth} It is known from an earlier study that $P_{3}$ increases due to the rotation of $P$ with increasing length along the [001] direction.~\cite{ederer,jang} Therefore, the observation of ${\alpha}_{31}{>}0$ for $H_{\rm{dc}}{>}0$ can be qualitatively understood as the result of an elongation of CFO/BFO along the [001] direction under $H_{\rm{dc}}{\parallel}[100]$ and the subsequent increase of $P_{3}$. The case of decreasing $P_{3}$ under $H_{\rm{dc}}{\parallel}[001]$ can also be understood in a similar way because the longitudinal magnetostriction ($\lambda_{33}$) in the CFO crystal was negative.

However, our MES data are seemingly inconsistent with the magnetostriction behavior of a bulk CFO at the quantitative level. The maximum to minimum value of $\alpha_{31}$ ($\Delta \alpha _{31}$) in Fig. 3(a), amounting to $\sim$260 ps/m ($\sim$120 mV/cm Oe), is about five times larger than that of $\alpha _{33}$ ($\Delta \alpha _{33}$). According to the magnetostriction of a Co$_{0.8}$Fe$_{2.2}$O$_{4}$ crystal with demagnetized domains (Fig. 3(c)), a slope of the $\lambda _{33}$ vs. $H$ curve is at least twice that of the $\lambda _{31}$ vs. $H$ curve.~\cite{bozorth} Upon assuming that the magnetostriction of CFO nanopillars follows a bulk behavior, these magnetostriction data predict that $\Delta \alpha _{33}$ should be at least twice of $\Delta \alpha _{31}$, which is in sharp contrast with the results in Fig. 3 (a) and (b).

Although there might exist several mechanisms to induce enhanced $\alpha _{31}$ as discussed in a recent anisotropic MES study using large $H_{\rm{ac}}$,~\cite{yanjap} one most decisive factor could be the preferred magnetic domains existing in the CFO nanopillars. The solid lines in Fig. 3(c) reproduce published $\lambda ^{\rm{s}}_{31}$ and $\lambda ^{\rm{s}}_{33}$ for a Co$_{0.8}$Fe$_{2.2}$O$_{4}$ crystal with a single magnetic domain along the [001] direction. The single magnetic domain was obtained through the magnetic annealing process, i.e. cooling under $H$ from high to room temperature.~\cite{bozorth} In this situation, applied $H{\parallel}[001]$ gave rise to the 180$^{\circ}$ domain wall motion, which resulted in negligible $\lambda ^{\rm{s}}_{33}$. In contrast, $H{\parallel}[100]$ resulted in a 90$^{\circ}$ domain wall motion so that it produced quite large $\lambda ^{\rm{s}}_{31}$. As we discussed above, the compressive, heteroepitaxial strain was a main source of enhanced magnetic anisotropy along the [001] direction in the CFO nanopillars. It is thus likely that the magnetic domains inside the CFO nanopillars have preferred alignment along the [001] direction, as illustrated in Fig. 3(f). If so, similar to the case of single magnetic domains, the CFO nanopillars are expected to have enhanced $\lambda _{31}$ and suppressed $\lambda _{33}$. As a result, as observed in Fig. 3 (a) and (b), the BFO-CFO nanostructure will give rise to a bigger $\alpha _{31}$ (smaller $\alpha _{33}$) than that expected based on the behavior of bulk CFO magnetostriction.

These results point to the possibility that the strain-induced ME coupling in the nanostructured film can be quite different from the macroscopic bulk composite. Application of a compressive, heteroepitaxial strain to the CFO nanopillars enables to achieve increased $\alpha _{31}$ to as high as $\sim$130 ps/m ($\sim$60 mV/cm Oe) at 6 kOe. Upon increasing $|u_{001}|$, as done in the BTO-CFO nanostructures with growth temperatures,~\cite{zhengapl} the $\alpha _{31}$ can be further optimized. In comparison, our previous study on a thin film made of NiFe$_{2}$O$_{4}$ nanoparticulates embedded in a PbZr$_{0.52}$Ti$_{0.48}$O$_{3}$ matrix [i.e., the (0-3) structure] showed maximum $|\alpha _{31}|{=}4$ mV/cm Oe ($\sim$4 ps/m) and $|\alpha _{33}|{=}16$ mV/cm Oe ($\sim$14 ps/m)~\cite{hryu}, which was clearly smaller than the maximum $\alpha _{31}{\sim}130$ ps/m found here. Therefore, our results strongly support that thin films with the (1-3) nanostructure have larger ME couplings than the other nanostructures [e.g., (0-3) structure].

In conclusion, we have determined the anisotropic MES of a 300 nm thick BiFeO$_{3}$-CoFe$_{2}$O$_{4}$ nanostructure. An enhancement was observed in the transverse configuration, which can be explained by the preferred alignment of magnetic domain resulting from the heteroepitaxial strain that is unique to the present (1-3) nanostructure. This investigation offers quantitative evidences that the nanoscale engineering of strain coupling is useful for the design of ME devices.

We appreciate discussions with T. W. Noh. This study was supported by the NRF, Korea through National Creative Research Initiatives, NRL (M10600000238) and Basic Science Research (2009-0083512) programs and by MOKE through the Fundamental R${\&}$D Program for Core Technology of Materials. YSO is supported by Seoul R${\&}$BD (10543). Y.H.C. is supported by the National Science Council, R.O.C. (NSC 98-2119-M-009-M016).

\end{document}